# Women Scientists Who Made Nuclear Astrophysics

◉Christine V. Hampton[1], ◉Maria Lugaro[2], Panagiota Papakonstantinou[3], P. Gina Isar[4], Birgitta Nordström[5], Nalan Özkan[6], Marialuisa Aliotta[7], Aleksandra Ćiprijanović[8], Sanjana Curtis[9], Marcella Di Criscienzo[10], Jacqueline den Hartogh[2], Andreea S. Font[11], Anu Kankainen[12], Chiaki Kobayashi[13], Claudia Lederer-Woods[7], Ewa Niemczura[14], Thomas Rauscher[13,15], Artemis Spyrou[16], Sophie Van Eck[17], Mariya Yavahchova[18], William Chantereau[11], Selma E. de Mink[19], Etienne Kaiser[20], Friedrich-Karl Thielemann[15,21], Claudia Travaglio[22], Aparna Venkatesan[23], Remo Collet[24]

[1] Hampton Consulting, LLC, Okemos, MI 48805, USA
`chrisvha@umich.edu`
[2] Konkoly Observatory, Hungarian Academy of Sciences, H-1121 Budapest, Hungary
`maria.lugaro@csfk.mta.hu`
[3] Institute for Basic Science, Rare Isotope Science Project, Daejeon 34047, South Korea
[4] Institute of Space Science, Bucharest-Magurele 077125, Romania
[5] Niels Bohr Institute, Juliane Maries vej 30, DK-2100 Copenhagen, Denmark
[6] Department of Physics, Kocaeli University, Umuttepe 41380, Kocaeli, Turkey
[7] SUPA, School of Physics and Astronomy, Univ. of Edinburgh, Edinburgh EH9 3FD, UK
[8] Department of Astronomy, Faculty of Mathematics, University of Belgrade, Serbia
[9] Department of Physics, North Carolina State University, Raleigh, NC 27606, USA
[10] INAF - Osservatorio di Roma, via Frascati 33, Monteporzio Catone, Rome, Italy
[11] Astrophysics Research Institute, Liverpool John Moores University, Liverpool, L3 5RF, UK
[12] Department of Physics, University of Jyväskylä, PO Box 35 (YFL), F1-40014, Finland
[13] Centre for Astrophysics Research, University of Hertfordshire, Hatfield AL 109AB, UK
[14] University of Wroclaw, Kopernika 11, PL-51-622 Wroclaw, Poland
[15] Department of Physics, University of Basel, 4056 Basel, Switzerland
[16] NSCL / JINA-CEE, Department of Physics and Astronomy, Michigan State University, USA
[17] Institut d'Astronomie et d'Astrophysique, Université libre de Bruxelles, Belgium
[18] Institute for Nuclear Research & Nuclear Energy, Bulgarian Academy of Sciences, Bulgaria
[19] Anton Pannekoek Institute for Astronomy, University of Amsterdam, The Netherlands
[20] Astrophysics Group, Keele University, ST5 5BG Keele, UK
[21] GSI Helmholtz Center for Heavy Ion Research, Darmstadt, Germany
[22] INFN-Turin, Astrophysical Observatory Turin, Italy
[23] Dept. of Physics and Astronomy, Univ. of San Francisco, San Francisco, CA 94117, USA
[24] Stellar Astrophysics Centre, Dept. of Physics and Astronomy, Aarhus University, Denmark

**Abstract.** Female role models reduce the impact on women of stereotype threat, i.e., of being at risk of conforming to a negative stereotype about one's social, gender, or racial group [1,2]. This can lead women scientists to underperform or to leave their scientific career because of negative stereotypes such as, not being as talented or as interested in science as men. Sadly, history rarely provides role models for women scientists; instead, it often renders these women invisible [3]. In response to this situation, we present a selection of twelve outstanding women who helped to develop nuclear astrophysics.

**Keywords:** Women Scientists, Nuclear Physics, Astrophysics, Radioactivity, Fission, Astronomy, Cosmology, Solar Studies, Historical

# 1 Introduction

Nuclear astrophysics is a melding of theoretical and experimental nuclear physics, observational astronomy, astrophysical modeling, and cosmological theory. It involves spectroscopic identifications, star classifications, prediction and discovery of stellar objects, construction of instrumentation, and chemical and physical interpretations. Women scientists have been an essential part of the development of these fields.

# 2 Twelve Scientists

**Marie Skłodowska Curie (1867-1934)** chose to investigate radiation phenomena in 1896 for her PhD, and in doing so, explained the theoretical basis of radioactivity; developed methods for isolating radioactive isotopes; and discovered the elements, Po and Ra [4]. Marie has the distinction of being the first female Nobel Laureate (Physics 1903; Chemistry 1911). Prof. Curie's outstanding achievements in Physics, Chemistry, Radiology, and Medicine; her humanitarian efforts in the throes of WWI; and her response to challenges will continue to inspire us for generations to come [5].

**Lise Meitner (1878-1968)** was the second woman at the University of Vienna to receive a doctorate in Physics and the first woman in Germany to become a full professor. Her most significant achievement was the theoretical explanation of nuclear fission [6]. She also discovered a number of radioactive isotopes together with Otto Hahn, with whom she collaborated for 30 years. Their discovery of Pa-231 was instrumental in establishing Protactinium as an element [7]. Prof. Meitner was nominated for the Nobel Prize 48 times (29 in Physics; 19 in Chemistry) [8].

**Ștefania Mărăcineanu (1882-1944),** after a teaching career in secondary schools in Romania, obtained a fellowship at the Radium Institute working with Marie Curie. In 1924, she defended her PhD at the Sorbonne on the half-life of Po [9]. She also researched the interaction of Po radiation with metals. With this work she may have introduced the 'philosophical concept' of artificial radioactivity [10]. After her PhD, Dr. Mărăcineanu worked on developing techniques for atmospheric nucleation reactions; then returned to Romania in 1930 to install their first Radioactivity Laboratory.

**Cecilia Payne Gaposchkin (1900-1979)** worked as a human-computer at the Harvard Observatory [11]. During her PhD, she made the discovery that the strength of stellar spectral lines depend not only on the stellar surface composition, but also on the degree of ionization at a given temperature. She was the first to conclude that hydrogen and helium are much more abundant in stars than all other elements [12]. Prof. Payne-Gaposchkin became the first female full-professor at Harvard's Faculty of Arts and Sciences and the first woman chair of a department at Harvard.

**Maria Goeppert Mayer (1906-1972).** Magic nucleon numbers, reflected in nuclear properties and in the observed solar abundances, had long puzzled physicists. In 1949, Maria devised a brilliant solution by coupling the nucleon spin with its orbital parameter [13]. She was hired for the Manhattan project, at the University of Chicago, and at Argonne National Laboratory. Prof. Mayer's work on magic numbers won her the Nobel Prize with Hans Jensen for their discoveries concerning nuclear shell structure.

**Toshiko Yuasa (1909-1980)** was Japan's first female nuclear physicist. She specialized in spectroscopy. In 1939 she won a French scholarship to work with Frédéric Joliot-Curie, resulting in a doctorate from the Collège de France on the continuous β-ray spectrum in artificial radioactive material. After WWII, Toshiko worked at Riken Nishina Center. She returned to France in 1949 to continue her nuclear research at CNRS. Her interests turned to reactions with synchrocyclotrons and in 1962, Prof. Yuasa earned a second doctorate from Kyoto University on the β decay of $^6$He [14].

**Georgeanne (Jan) R. Caughlan (1916-1994)** studied nuclear data of reactions important for stars. Her very first efforts to provide extensive compilations of nuclear reaction rates based on current experimental information resulted in some of the most famous papers in the field [15]. Jan's career followed a very nontraditional path. After receiving a Physics degree, she dedicated herself to raising her five children. Prof. Caughlan earned her PhD at the age of 48; became Professor of Physics at the age of 58; and then became Acting Dean of Graduate Studies at Montana State University.

**Edith Alice Müller (1918-1995)** obtained her PhD in solar physics and worked on the observation and theory of the solar atmosphere. With her collaborators, Goldberg and Aller she published an extremely influential paper on the elements in the solar atmosphere which remained the gold standard for 20 years [16]. Edith was the first woman to be appointed General Secretary of the IAU. A named award in Prof. Müller's honor will be granted in 2018 to an outstanding PhD thesis in Switzerland.

**E. Margaret Peachey Burbidge (b. 1919)** has played a central role in shaping nuclear astrophysics. Her early research focused on chemical abundances in stars. Her landmark 1957 paper [17] thrust the theory of stellar nucleosynthesis into the scientific spotlight. For her pioneering research, Margaret received 12 honorary degrees; was elected a Fellow of the Royal Society of London; and held many leadership positions, including becoming the first female president of the AAS. Prof. Burbidge is currently Professor Emeritus at the University of California, San Diego.

**Erika Helga Ruth Böhm-Vitense (1923-2017)** was the first scientist to accurately describe convective mixing in stellar interiors using a prescription that has been widely adopted for half a century now in all stellar evolutionary codes. Her 1958 paper [18] is a crucial contribution to "mixing-length" theory. She combined theory and observations in optical studies of a large variety of objects from helium stars, to super giants, to open clusters. Prof. Böhm-Vitense received the Annie Jump Cannon Prize from AAS and the Karl Schwarzschild Medal from the Astronomische Gesellschaft.

**Dilhan Ezer Eryurt (1926-2012)** is regarded as the mother of Astronomy in Turkey; her life dedicated to science created a tremendous legacy. After completing her doctorate, she worked at Indiana University, NASA Goddard Space Flight Center, and the University of California. Her work revealed a striking fact about the Sun: it was much brighter and warmer in the past than it is today [19]. Prof. Eryurt received the 1969 Apollo Achievement Award; organized the first National Astronomy Congress in Turkey; founded the Astrophysics branch in Physics at the Middle East Technical University; and became Chair of the Department and Dean of the Faculty.

**Beatrice Muriel Tinsley (1941-1981)** was a true pioneer of the chemical evolution of galaxies. In her 1980 review article [20], we find her brilliant explanations of the modeling of galaxies and her predictions that are still pertinent today. With her PhD dissertation awarded by the University of Texas in 1967, she started her journey into achieving international fame as a cosmologist. Her work was considered revolutionary with the discovery that the Universe was in a state of infinite expansion [21]. Prof. Tinsley was the first woman at Yale to advance to Professor of Astronomy.

## 3   Summary

We present role models for young scholars and encourage them to explore nuclear astrophysics as a potential career path. Our intent is also to remind the scientific community and to inform the general public about the significant role women have played and continue to play in the development of Nuclear Astrophysics. Our poster contains additional information and will be freely available for download from the ChETEC website (www.ChETEC.eu).

**Acknowledgements:** This work is supported by ChETEC Action (CA16117) which is supported by COST (www.cost.eu). COST (European Cooperation in Science and Technology) is a funding agency for research and innovation networks. The partial support of RISP/ IBS, funded by the Ministry of Science, ICT and Future Planning and the National Research Foundation of Korea (2013M7A1A1075764) is acknowledged. The work of PGI was supported by the Romanian National Authority for Scientific Research and Innovation, CNCS/CCCDI-UEFISCDI, project numbers PN-III-P2-2.1-PED-2016-0339 and PN-III-P1-1-2-PCCDI-2017-0839 within PNCDI III.